\documentclass[aps,pra,reprint,superscriptaddress,showpacs,longbibliography]{revtex4-1}
\usepackage{amssymb,amsthm,amsmath,bbm,bbold,graphicx,epsfig,threeparttable,color}
\usepackage[colorlinks=true,linkcolor=blue]{hyperref}
\usepackage{ulem,enumerate}

\theoremstyle{definition}

\begin{document}

\title{Thermodynamic description of non-Markovian information flux of nonequilibrium open quantum systems}
\author{Hong-Bin Chen}
\email{hongbinchen@phys.ncku.edu.tw}
\affiliation{Department of Physics, National Cheng Kung University, Tainan 70101, Taiwan}
\author{Guang-Yin Chen}
\affiliation{Department of Physics, National Chung Hsing University, Taichung 40227, Taiwan}
\author{Yueh-Nan Chen}
\email{yuehnan@mail.ncku.edu.tw}
\affiliation{Department of Physics, National Cheng Kung University, Tainan 70101, Taiwan}
\affiliation{Physics Division, National Center for Theoretical Sciences, Hsinchu 30013, Taiwan}
\date{\today}

\begin{abstract}
One of the fundamental issues in the field of open quantum systems is the classification and quantification of non-Markovianity. In the contest of quantity-based measures
of non-Markovianity, the intuition of non-Markovianity in terms of information backflow is widely discussed. However, it is not easy to characterize the information flux
for a given system state and show its connection to non-Markovianity. Here, by using the concepts from thermodynamics and information theory, we discuss a potential
definition of information flux of an open quantum system, valid for static environments. We present a simple protocol to show how a system attempts to share information
with its environment and how it builds up system-environment correlations. We also show that the information returned from the correlations characterizes the
non-Markovianity and a hierarchy of indivisibility of the system dynamics.
\end{abstract}

\maketitle

\section{INTRODUCTION}

A detailed understanding of how a quantum system interacts with an environment is important for a wide variety of fields \cite{leggett_rev_rmp_1987,breuer_textbook,weiss_textbook,neill_nat_phy_2013,hongbin_hami_ensem_simu_arxiv_2017,ines_non_mark_review_rmp_2017}.
One of the fundamental issues in this topic is a complete description of non-Markovian effects, i.e., memory properties of the system-environment
interaction which cannot be captured by the conventional Born-Markov approximation. For example, many efforts have been devoted to the quantification of
non-Markovianity in open quantum systems \cite{ines_non_mark_review_rmp_2017,plenio_non_mark_review_rep_prog_phys_2014,breuer_non_mark_review_rmp_2016}.
Several practical measures of non-Markovianity have been proposed, typically based on the expected monotonicity of certain quantities under completely
positive and trace-preserving (CPTP) maps
\cite{BLP_measure_prl_2009,RHP_measure_prl_2010,LFS_measure_pra_2012,LPP_measure_pra_2013,BCM_measure_sci_rep_2014,acc_info_measure_prl_2014,acc_info_measure_pra_2014,shinlian_tsw_measure_prl_2016}.
The central idea is that when these quantities show monotonicity, as a function of time, the system dynamics can be classified as Markovian. In contrast,
whenever these quantities violate monotonicity, the dynamics are classified as non-Markovian and the map which describes the dynamics is said to be
indivisible \cite{RHP_measure_prl_2010,wolf_cmp_2008,chruscinski_k_divi_prl_2014,hongbin_k_divi_pra_2015} or strong non-Markovian \cite{bernardes_experiment_k_divi_scirep_2015}.
A measure of non-Markovianity can thus be constructed according to the overall nonmonotonic part of these quantities.

One physical interpretation of the monotonicity of such quantities under CPTP maps can be gained from the so-called data processing theorem
\cite{man_qi_text_book,francesco_prl_2014,francesco_pra_2016}. This says that, for a Markovian process, information continuously dissipates out of the system. Therefore,
any retrieved knowledge on the system state from the environment characterizes the non-Markovianity of the process. For instance, in the non-Markovianity measure proposed
by Breuer, Lane, and Piilo (BLP) \cite{BLP_measure_prl_2009}, the authors focus on the trace distance of a pair of arbitrary initial states and show that the revival of
trace distance witnesses a backflow of information, which increases the distinguishability of the state pair and, consequently, characterizes the non-Markovianity.

However, it is often not easy to characterize the ``information'' for a given system undergoing a dynamical process without referring to any ancillary degrees of freedom.
Moreover, existing quantity-based measures, while each having various benefits, tend to show discrepancies
\cite{hongbin_k_divi_pra_2015,apollaro_comparison_pra_2014,addis_comparison_pra_2014} between each other. Consequently, we expect that more rigorously characterizing
non-Markovianity in terms of information flux will assist in concretely defining the nature of non-Markovianity, and in developing new measures in the future.

On the other hand, information theory, and its interplay with
thermodynamics \cite{bennett_compu_td_review_ijtp_1982,plenio_review_contemp_phys_2001,maruyama_review_rmp_2009,toyabe_info_energy_conv_nat_phys_2010,
parrondo_review_nat_phys_2015,faist_work_cost_info_nat_comm_2015,philipp_qi_qtd_prx_2017} has helped reveal the nature of information not as an abstraction, but as a
physical resource. In this work we discuss how the language of thermodynamics and information theory is used to explicitly define the information flux through an open
system, and in turn the non-Markovianity.

To this end, we revisit the thermodynamic task of work extraction \cite{allahverdyan_work_ext_epl_2004,allahverdyan_work_ext_epl_2011,alicki_td_qi_op_sys_info_dyn_2004,aberg_work_ext_nat_comm_2013,skrzypczyk_work_ext_nat_comm_2014}
and the thermodynamic quantity, entropy production \cite{spohn_ent_prod_jmp_1978,breuer_ent_prod_pra_2003,esposito_ent_prod_corr_njp_2010,deffner_ent_prod_prl_2011,esposito_ent_prod_epl_2011},
in nonequilibrium situations. First, we define the information flux via the negative entropy production rate, and show that the system tends to share the outgoing
information with its environment and establish system-environment correlations. For a convincing demonstration of these definitions, we then discuss a protocol based on
a thermodynamic process involving a two-level system with resonant components of a reservoir \cite{skrzypczyk_work_ext_nat_comm_2014,brunner_virt_qubit_virt_temp_pre_2012},
which reaffirms our main results.

{To describe our definition within the framework of open system, we will then consider how these quantities can be defined in terms of Lindblad superoperator prescription,
and use this to discuss the non-Markovianity of a qubit pair coupled with each other via a controlled-NOT (CNOT) gate. This will help us see how information is exchanged
in terms of system-environment correlations during a dynamical process, and how the information flux can be used to fully characterize the hierarchy of indivisibility and
non-Markovianity. We will also discuss why the BLP measure \cite{BLP_measure_prl_2009} has difficulty in capturing all of the information backflow in this example.

\section{WORK EXTRACTION AND INFORMATION IN A NONEQUILIBRIUM SYSTEM}

Before explicitly defining information flux, we must understand how to quantify the amount of information, $I^\mathrm{neq}(\rho)$, encoded in terms of a state
configuration $\rho$ out of equilibrium. Given the important link between the task of work extraction and information theory, as appears in the examples of
Maxwell's demon \cite{callen_thermo_text_book}, the Szil\'{a}rd engine \cite{szilard_z_phys_1929}, and Landauer's erasure principle \cite{landauer_principle_ibmjrd_1961},
it is becoming more common to consider the nature of information as physical. For example, in the Maxwell's demon example, the demon operates a Szil\'{a}rd engine,
consisting of a single ideal gas molecule and a chamber divided into two sides with equal volume. The demon is capable of accessing the initial position of the molecule.
By consuming this knowledge, the demon can extract an average amount of work $W^\mathrm{ext}=k_\mathrm{B}T\mathrm{ln}2$ from a heat reservoir at temperature $T$, where
$k_\mathrm{B}$ is the Boltzmann constant.

In the general case, for a system with non-trivial Hamiltonian $\widehat{H}_\mathrm{sys}$, then the maximal amount of average extractable work by using the system in an
initial state $\rho$, before it equilibrates with a reservoir at temperature $T$, is given by the change in free energy
\cite{alicki_td_qi_op_sys_info_dyn_2004,aberg_work_ext_nat_comm_2013,skrzypczyk_work_ext_nat_comm_2014,esposito_ent_prod_epl_2011}
\begin{equation}
W^\mathrm{ext}=F(\rho)-F(\rho^\mathrm{eq})=k_\mathrm{B}TS\left(\rho\Vert\rho^{\mathrm{eq}}\right),
\end{equation}
where $F(\rho)=\langle\widehat{H}_\mathrm{sys}\rangle-k_\mathrm{B}TS(\rho)$, the Helmholtz free energy, is one of the most fundamental quantities in thermodynamics,
$F(\rho^\mathrm{eq})=-k_\mathrm{B}T\mathrm{ln}Z$ is the free energy at thermal equilibrium, $S(\rho)=-\mathrm{Tr}\rho\mathrm{ln}\rho$ is the von Neumann entropy,
$S\left(\rho_1\Vert\rho_2\right)=\mathrm{Tr}\rho_1\left(\mathrm{ln}\rho_1-\mathrm{ln}\rho_2\right)$ is the relative entropy (Kullback-Leibler divergence),
$\rho^\mathrm{eq}=\exp\left[-\widehat{H}_\mathrm{sys}/k_\mathrm{B}T\right]/Z$, and $Z=\mathrm{Tr}\exp\left[-\widehat{H}_\mathrm{sys}/k_\mathrm{B}T\right]$ is the partition
function, respectively.

The significance of a general Szil\'{a}rd engine is that it conjoins thermodynamics and information theory. It shows the usefulness of information for performing some
thermodynamic tasks. Motivated by the task of work extraction, one can therefore quantify the amount of information encoded in a state configuration $\rho$ with respect
to its thermal equilibrium $\rho^\mathrm{eq}$ via
\begin{equation}
I^\mathrm{neq}(\rho)=S\left(\rho\Vert\rho^\mathrm{eq}\right).
\label{eq_sys_equi_rel_ent}
\end{equation}

This definition is different to the Shannon entropy for a probability distribution or von Neumann entropy for a quantum state generically adopted in standard information
theory. Intuitively, whenever a system is more pure, it is usually more useful for extracting work. But it possesses less von Neumann entropy since it is less uncertain
(i.e., requires less information to encode). Here, inspired by the non-Markovianity measure theory, we consider the ``usefulness" or ``purity" of a state as a definition
of information rather than the uncertainty of a state.

\section{INFORMATION FLUX THROUGH OPEN SYSTEMS}

\subsection{Definition of information flux}

When a system undergoes a dynamical process, the change in entropy of the system originates from two sources
\begin{equation}
\Delta S_\mathrm{sys}=\Delta S^\mathrm{rev}+\Delta S^\mathrm{irr},
\end{equation}
where $\Delta S^{\mathrm{rev}}=\overline{\Delta}Q/k_\mathrm{B}T$ is the reversible entropy change arising from exchanging heat $\overline{\Delta}Q$ with the environment, and
the irreversible contribution $\Delta S^\mathrm{irr}=\Delta S_\mathrm{sys}-\overline{\Delta}Q/k_\mathrm{B}T$ is referred to as the entropy production. The heat exchange is
defined as
\begin{equation}
\overline{\Delta}Q=\int_0^t\mathrm{Tr}\left[\widehat{H}_\mathrm{sys}(\tau)\frac{\partial}{\partial \tau}\rho(\tau)\right]d\tau,
\end{equation}
which is positive if heat is flowing into the system and negative if reversed. The overline reminds readers that this quantity is path dependent, rather than a state
function. The system Hamiltonian $\widehat{H}_\mathrm{sys}(t)$ can be time dependent in general. The sources of time dependence may come from external driving or the
interaction with environment.

Irreversibility is an ubiquitous phenomenon in nature. Historically, this was conceived as an empirical axiom and stated in terms of the second law of thermodynamics. The
positivity of entropy production $\Delta S^{\mathrm{irr}}\geq 0$ is called the Clausius inequality and is one of the various ways of expressing the second law. Therefore,
$\Delta S^{\mathrm{irr}}$ is customarily said to be irreversible. Inspired by the quantity-based non-Markovianity measures described in the Introduction, one may pose the
question of whether the positivity of entropy production is also promising for constructing a practical measure of non-Markovianity. One may also ask the following: how does the entropy
production characterize the information flux out of the system and how does it relate to the non-Markovianity of a dynamical process?

One can expect that the information flowing out of a system should be either transferred into the environment or contained in the system-environment correlations (e.g., in
the form of quantum entanglement). The former is encoded in the form of heat transfer, namely the reversible entropy change $\Delta S^\mathrm{rev}$. Hence its nature is more
``energetic.'' We are particularly interested in the latter, which is associated to the irreversible entropy production $\Delta S^\mathrm{irr}$ and has a more
``informational'' meaning. This intuition is schematically shown in Fig.~\ref{fig_entropy_change_route} and will become clear in the following. Since the system-environment
correlation is so fragile and suffers damage from the environmental fluctuations, it is therefore responsible for the irreversibility associated to the entropy production.

Inspired by the above intuition, here we define, for a sluggish or static environment, the total information flux $\mathcal{F}$ through the system is equal to the negative
entropy production rate, i.e.,
\begin{equation}
\mathcal{F}=-\frac{\partial S^\mathrm{irr}}{\partial t}.
\label{eq_info_flux_definition}
\end{equation}
We emphasize that, in principle, the entropy production rate $\partial_t S^\mathrm{irr}$ can be calculated for any general case with vigorous environments. Nevertheless, its
capability of characterizing the information flux becomes ambiguous in such general cases since we quantify the amount of information in $\rho$ with $I^\mathrm{neq}(\rho)$ in
Eq.~(\ref{eq_sys_equi_rel_ent}), which is based on the task of work extraction from a static reservoir. Besides, some of our following arguments rely on this hypothesis as well.
We will argue that, under the hypothesis of static environment, Eq.~(\ref{eq_info_flux_definition}) can be related to the system-environment correlations and the non-Markovianity
of open quantum systems, and demonstrate several protocols and examples, which reaffirm our definition.

\begin{figure}[th]
\includegraphics[width=\columnwidth]{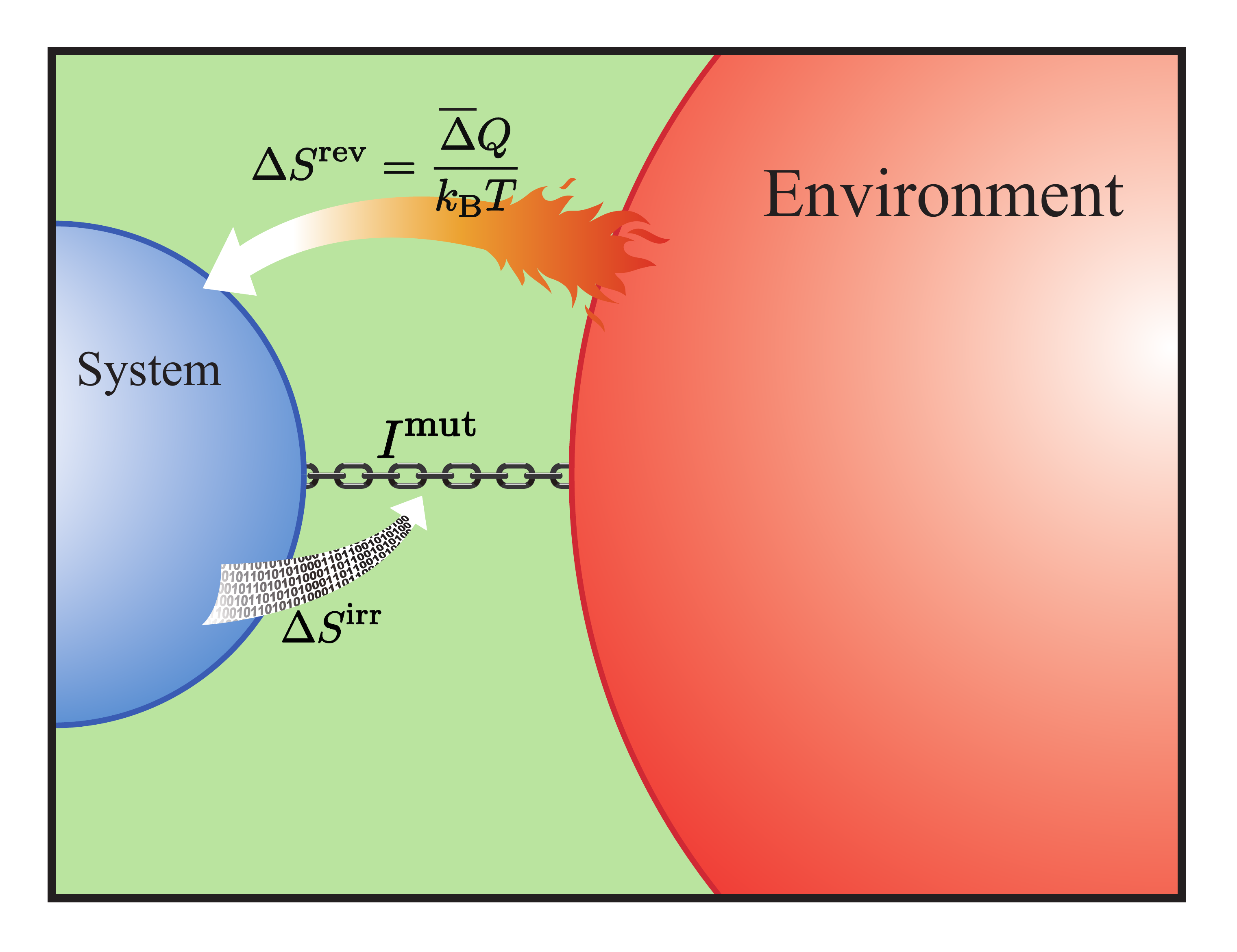}
\caption{Change in system entropy is partitioned into two terms. The reversible entropy change $\Delta S^\mathrm{rev}=\overline{\Delta}Q/k_\mathrm{B}T$ arises from
exchanging heat with the environment, while the irreversible contribution $\Delta S^\mathrm{irr}$, referred to as the entropy production, characterizes the information flux
out of the system. The system attempts to share this outflowing information with the environment and establishes system-environment correlations $I^\mathrm{mut}$.}
\label{fig_entropy_change_route}
\end{figure}

\subsection{Static environment hypothesis}

In the task of work extraction, the reservoir is considered to be static in the sense that the perturbation given by a finite dimensional system is negligibly small and will
relax in a time scale much shorter than the characteristic time of system dynamics. Hence reservoir's population is assumed to be fixed and obeys the Boltzmann distribution.

More precisely, it is assumed that the environment deviates from thermal equilibrium by a small variation during a dynamical process, i.e.,
$\rho_\mathrm{env}(t)=\rho_\mathrm{env}^\mathrm{eq}+\delta\rho(t)$ with $\mathrm{Tr}\delta\rho(t)=0$. As pointed out in Ref.~\cite{philipp_qi_qtd_prx_2017}, the information
stored in the environmental configuration, in analog to Eq.~(\ref{eq_sys_equi_rel_ent}), is expressed as
\begin{eqnarray}
I_\mathrm{env}^\mathrm{neq}(t)&=&S\left(\rho_\mathrm{env}(t)\Vert\rho_\mathrm{env}^\mathrm{eq}\right) \nonumber\\
&=&-\frac{\overline{\Delta}Q}{k_\mathrm{B}T}-\Delta S_\mathrm{env} \nonumber\\
&=&\frac{1}{2}\mathrm{Tr}\left[\left(\rho_\mathrm{env}^\mathrm{eq}\right)^{-1}\delta\rho^2\right]+\mathcal{O}\left(\delta\rho^3\right),
\label{eq_rel_ent_env_eqenv}
\end{eqnarray}
which becomes vanishingly small as $\delta\rho\rightarrow0$. Therefore, the entropy change of the environment is solely described by the amount of heat flowing into the system
\begin{equation}
\Delta S_\mathrm{env}=-\frac{\overline{\Delta}Q}{k_\mathrm{B}T}.
\label{eq_diff_entr_env}
\end{equation}

In this work, we may slightly release the assumption. Namely, we solely require $\rho_\mathrm{env}=\mathrm{Tr}_\mathrm{sys}\rho_\mathrm{tot}(t)$ to be time independent, but
not necessarily homogeneously thermalized. This also implicitly requires that the environment Hamiltonian is constant in time. This assumption is weaker than the conventional
Born approximation, which explicitly eliminates all system-environment correlations. Additionally, we stress that even though the environment is assumed to be static, the
system and the environment can still build significant correlations during evolution \cite{hongbin_hami_ensem_simu_arxiv_2017}, and the system dynamics can exhibit a
non-Markovian nature if it contacts to a structured environment with sufficient long correlation times \cite{hongbin_3sbm_scirep_2015}, even though the environment is
thermalized. This justifies the significance of our work.

\subsection{Information exchange of an open system}

Considering a system undergoing nonequilibrium dynamics with a time-dependent Hamiltonian $\widehat{H}_\mathrm{sys}(t)$. We define an ``instantaneous'' equilibrium
$\rho^\mathrm{eq}(t)=\exp\left[-\widehat{H}_\mathrm{sys}(t)/k_\mathrm{B}T\right]/Z_t$ at each time instance in a similar manner to the static case. The system starts from a
nonequilibrium initial state $\rho (0)$ and evolves to another nonequilibrium state $\rho(t)$ at a later time $t$.

As shown in Refs.~\cite{deffner_ent_prod_prl_2011,esposito_ent_prod_epl_2011}, the change in the information of the system during the dynamical process is given by
\begin{equation}
\Delta I^\mathrm{neq}=I_t^\mathrm{neq}-I_0^\mathrm{neq}=-\Delta S^\mathrm{irr}+\frac{\overline{\Delta}W^\mathrm{irr}}{k_\mathrm{B}T}.
\label{eq_diff_neq_info}
\end{equation}
The two components on the right-hand side of Eq.~(\ref{eq_diff_neq_info}) have their own individual physical interpretations. The first term, $-\Delta S^\mathrm{irr}$, denotes
the contribution to the change in the information caused by state transformation, and the minus sign reflects that the information flowing out of the system gives rise to a
reduction of the residual information in the system. This means that the entropy production does characterize certain information lost in the system and supports our definition
in Eq.~(\ref{eq_info_flux_definition}). The second term $\overline{\Delta}W^\mathrm{irr}=\overline{\Delta}W-k_\mathrm{B}T \left(-\mathrm{ln}Z_t+\mathrm{ln}Z_0\right)$ is the
irreversible work \cite{esposito_ent_prod_epl_2011} and $\overline{\Delta}W=\int_0^t\mathrm{Tr}\left[\rho(\tau)\partial_\tau\widehat{H}_\mathrm{sys}(\tau)\right]d\tau$ is the
work performed on the system. Therefore, the irreversible work accounts for the contribution arising from the time variation of Hamiltonian. It is zero for the case of
constant Hamiltonian.

Consequently in our definition~(\ref{eq_info_flux_definition}), we only take the entropy production rate into account and ignore the contribution by irreversible work since the
entropy production rate quantifies the time-varying rate of information in the system caused by state transformation. In particular, given a dynamical process, one is usually
interested in state transition and may not clear how the Hamiltonian evolves.

\subsection{Geometric interpretation}

A heuristic geometric interpretation of Eq.~(\ref{eq_diff_neq_info}) is sketched in Fig.~\ref{fig_bloch_sphere}. The state space of the system forms a subset of positive
semidefinite operators with unit trace in a C$^{\ast}$ algebra of linear operators on the $n$-dimensional Hilbert space $\mathcal{H}_n$. For simplicity, we schematically
depict it as a Bloch sphere. As the system  Hamiltonian is time varying, the corresponding instantaneous eigenbasis is also time varying. Consider the diagonalized system
Hamiltonian in its corresponding eigenbasis; it can be expressed as a linear combination in the Cartan subalgebra of $\mathfrak{u}(n)$ and the component in $\mathfrak{su}(n)$
effectively defines a rotating ``$z$ axis'' of the Bloch sphere.

The instantaneous equilibrium states are always on the rotating $z$ axis and denoted as the red dots in Fig.~\ref{fig_bloch_sphere}. The nonequilibrium system states are
denoted by the black dots and the dynamics is represented by the black trajectory. The information $I^{\mathrm{neq}}$ (blue dashed line) can then be considered as the
``distance'' connecting the system state and the corresponding instantaneous equilibrium. As time proceeds, $I^{\mathrm{neq}}$ varies due to its two ends moving in the Bloch
sphere. Accordingly, the variation in $I^{\mathrm{neq}}$ consists of two contributions separately associated to the state transformation and the time-varying Hamiltonian, as
shown in Eq.~(\ref{eq_diff_neq_info}).

\begin{figure}[th]
\includegraphics[width=\columnwidth]{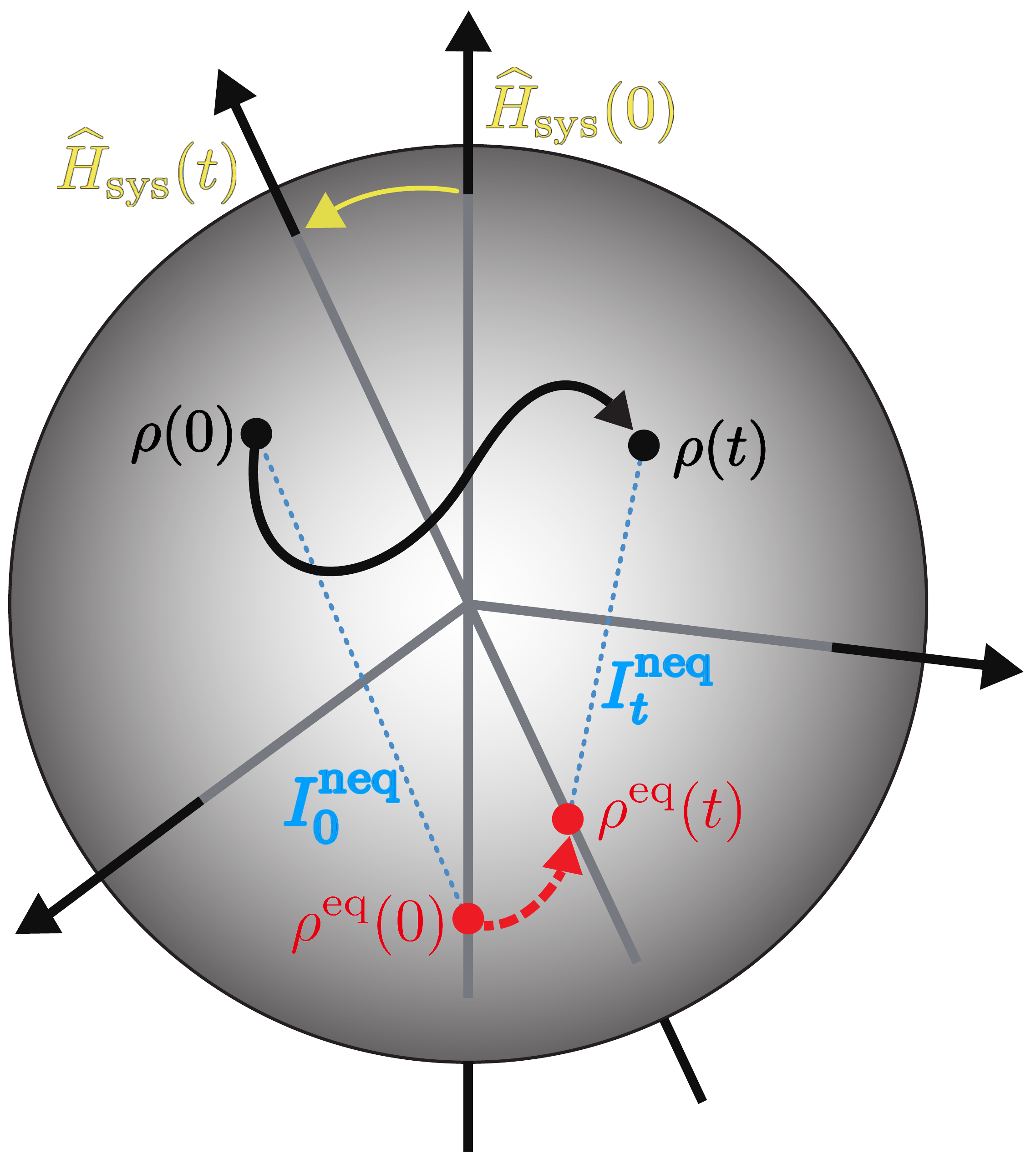}
\caption{We depict the state space as a simplified Bloch sphere. The time-varying Hamiltonian defines a rotating $z$ axis of the Bloch sphere. The instantaneous equilibrium
states always lie on the $z$ axis and are denoted by red dots. The dynamics of the system is represented by the black curve inside the Bloch sphere. The amount of
information $I^\mathrm{neq}(\rho)=S(\rho\Vert\rho^\mathrm{eq})$ is the length of the blue dashed straight line connecting the system state and the instantaneous equilibrium
state. Both state transformation and the time variations of the Hamiltonian give rise to the change in the information $I^\mathrm{neq}$, as shown in
Eq.~(\ref{eq_diff_neq_info}).}
\label{fig_bloch_sphere}
\end{figure}

\subsection{System-environment correlations}

Our second finding is that the system attempts to share outflowing information with the environment and establish system-environment correlations. A straightforward way to
visualize this is to consider the system and environment in totality as a closed system such that the total state $\rho_\mathrm{tot}(t)$ evolves unitarily without a change
in the total entropy. The bipartite mutual information, $I^\mathrm{mut}=S_\mathrm{sys}+S_\mathrm{env}-S_\mathrm{tot}$, quantifies the amount of information shared between the
two parties. In a closed total system the rate of change of the mutual information consists of the change in entropy of the system and the environment, i.e.,
$\partial_t I_t^\mathrm{mut}=\partial_t S_\mathrm{sys}+\partial_t S_\mathrm{env}$.

Assuming that the environment is static and kept thermalized at temperature $T$, then taking the time derivative form of Eq.~(\ref{eq_diff_entr_env}) leads to one of our main
results that the change in the mutual information comes from the information flowing through the system:
\begin{equation}
\frac{\partial I_t^\mathrm{mut}}{\partial t}=\frac{\partial S_\mathrm{sys}}{\partial t}-\frac{1}{k_\mathrm{B}T}\frac{\partial Q}{\partial t}
=-\mathcal{F}.
\label{eq_rat_mut_info_neg_info_flux}
\end{equation}
Namely, the information contained in the system-environment correlations is offered by the system \textit{per se}.

For a more precise consideration, suppose that the initial total state is a direct product of system and environment. In the beginning, we neither require the environment to
be thermalized nor static. One can show that, in a similar manner to Ref.~\cite{esposito_ent_prod_corr_njp_2010}, the entropy production can be expressed in terms of relative
entropy:
\begin{eqnarray}
\Delta S^\mathrm{irr}&=&S\left(\rho_\mathrm{tot}(t)\Vert\rho(t)\otimes\rho_\mathrm{env}^\mathrm{eq}\right)
-S\left(\rho_\mathrm{env}(0)\Vert\rho_\mathrm{env}^\mathrm{eq}\right)\nonumber \\
&=&I^\mathrm{mut}(t)+I_\mathrm{env}^\mathrm{neq}(t)-I_\mathrm{env}^\mathrm{neq}(0).
\label{eq_entr_pro_rel_entr}
\end{eqnarray}
More details of Eq.~(\ref{eq_entr_pro_rel_entr}) are shown in Appendix~\ref{app_derivation_eq_entr_pro_rel_entr}. Its meaning states that the entropy production of the system
not only quantifies the amount of mutual information, but also contains the information change caused by the environmental state transition. Finally, if we further assume that
the environment is static [i.e., $I_\mathrm{env}^\mathrm{neq}(t)=I_\mathrm{env}^\mathrm{neq}(0)$], it reduces to system-environment correlations exclusively:
\begin{equation}
\Delta S^\mathrm{irr}=I^\mathrm{mut}(t).
\end{equation}
This supports our intuition shown in Fig.~\ref{fig_entropy_change_route}. And then taking time derivative form immediately recovers our main result in
Eq.~(\ref{eq_rat_mut_info_neg_info_flux}).

\section{PROTOCOL}

Now we present a simple protocol (Fig.~\ref{fig_protocol}) to explicitly demonstrate Eq.~(\ref{eq_rat_mut_info_neg_info_flux}). We consider a two-level system as the
``system" in our protocol, with a nontrivial Hamiltonian $\widehat{H}_\mathrm{sys}=E_a|a\rangle\langle a|+E_b|b\rangle\langle b|$, where $E_{a}>E_{b}$. The initial state of
the system is given by $\rho_\mathrm{sys}=p_a|a\rangle\langle a|+p_b|b\rangle\langle b|$ with $p_a+p_b=1$. Although here we only consider a simplified model without initial
coherence, this restriction can be relaxed and generalized to that with initial coherence straightforwardly.

In this protocol the environment is assumed to be a huge reservoir in the sense that we can freely and repeatedly pick one copy of a virtual or ancillary two-level-system
(or qubit) \cite{skrzypczyk_work_ext_nat_comm_2014,brunner_virt_qubit_virt_temp_pre_2012}, which is on resonance with the real system, out of the environment, in each single
run of the protocol. Suppose that the two levels of the virtual qubit are labeled as $1$ and $0$ with $E_1>E_0$; then the state of the virtual qubit can be expressed as
\begin{equation}
\rho_\mathrm{vir}=\frac{q_1}{Z}|1\rangle\langle 1|+\frac{q_0}{Z}|0\rangle\langle 0|,
\end{equation}
where $q_1/q_0=\exp\left[-(E_a-E_b)/k_\mathrm{B}T\right]$ and $Z$ is the partition function of the environment.

To describe the ``thermal contact'' microscopically and in a quantum mechanical regime, we consider the interaction Hamiltonian
\begin{equation}
\widehat{H}_\mathrm{int}=\hbar\gamma\left(|b\rangle\langle a|\otimes|1\rangle\langle 0|+|a\rangle\langle b|\otimes|0\rangle\langle 1|\right).
\label{eq_prot_int_hami}
\end{equation}
The time evolution of the total system is then governed by the unitary
operator $\widehat{U}_t=\exp\left[-i\left(\widehat{H}_\mathrm{sys}+\widehat{H}_\mathrm{env}+\widehat{H}_\mathrm{int}\right)t/\hbar\right]$.

\begin{figure}[th]
\includegraphics[width=\columnwidth]{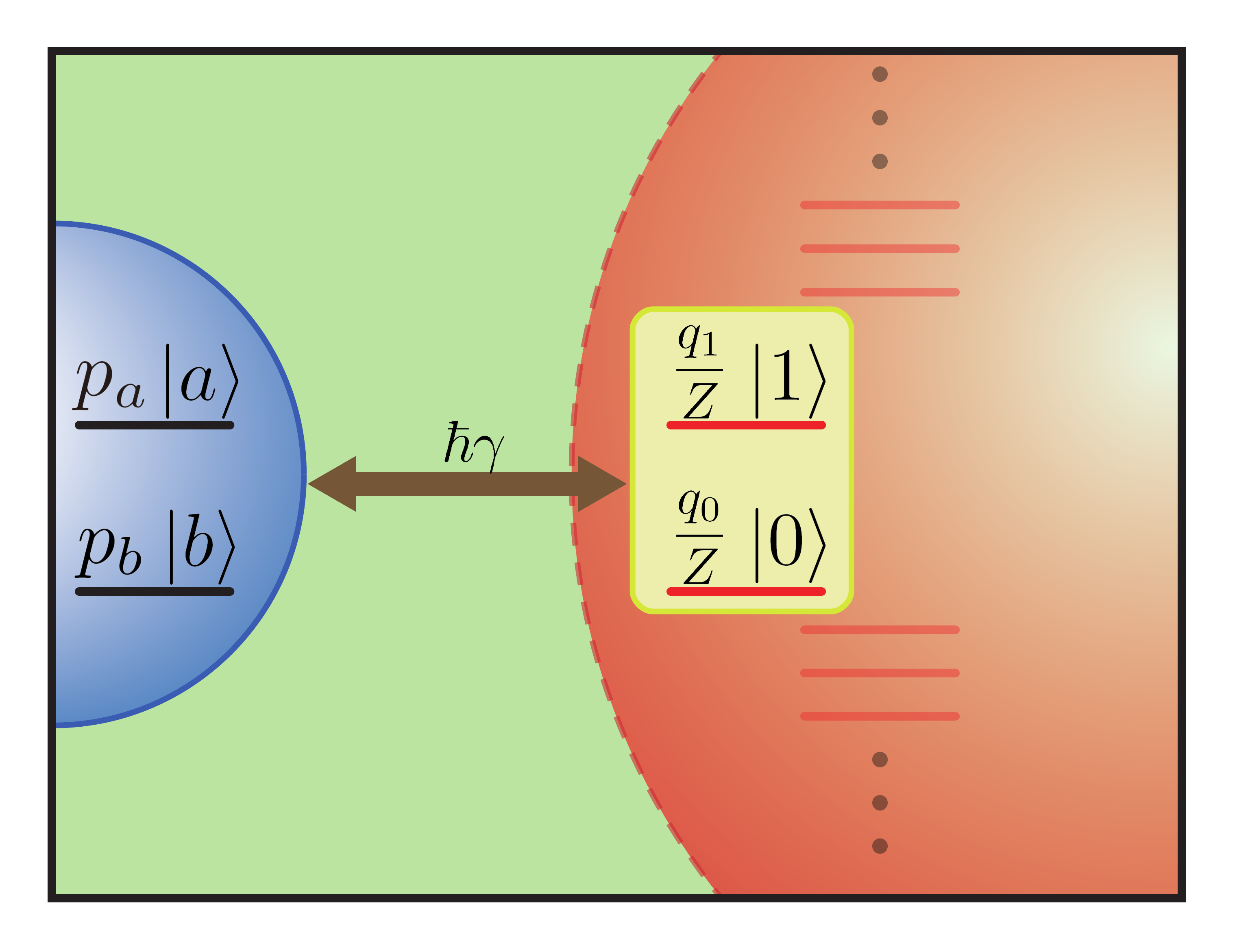}
\caption{We consider a simple protocol showing how the system-environment correlations are established by sharing the outflowing information. The system is modeled as a
two-level system with initial state $\protect\rho_\mathrm{sys}=p_{a}|a\rangle \langle a|+p_{b}|b\rangle \langle b|$. The environment is in thermal equilibrium at temperature
$T$ and we can repeatedly pick one virtual qubit $Z^{-1}(q_{1}|1\rangle \langle 1|+q_{0}|0\rangle \langle 0|)$ out of the environment in each single run of the protocol.}
\label{fig_protocol}
\end{figure}

The first stage of the protocol in each run is an infinitesimal evolution
\begin{equation}
\rho_\mathrm{tot}(0)\rightarrow\rho_\mathrm{tot}(\delta t)=\widehat{U}_{\delta t}\rho_\mathrm{tot}(0)\widehat{U}_{\delta t}^\dagger.
\end{equation}
The initial state $\rho_\mathrm{tot}(0)$ is a direct product of the system and the environment. After an infinitesimal evolution, heat $\overline{\delta}Q$ and information
$-\delta S^\mathrm{irr}$ are exchanged between the real system and the virtual qubit. Moreover, the correlation $\delta I^\mathrm{mut}$ is also established during the
infinitesimal evolution. In the second stage, we erase the correlations established in the first stage and obtain the reduced state of the system and the environment. We are
now able to calculate the heat and the amount of correlations induced by the infinitesimal evolution in the first stage. We finally discard the exhausted virtual qubit back
into the environment and again pick a new virtual qubit from the environment. Once again, we are ready for the next run of the protocol. Following
Ref.~\cite{modi_quan_corr_prl_2010}, the bipartite correlation can be quantified by the relative entropy
$\delta I^\mathrm{mut}=S\left(\rho_\mathrm{tot}(\delta t)\Vert\rho_\mathrm{sys}(\delta t)\otimes\rho_\mathrm{env}(\delta t)\right)$. Finally, we can conclude that
\begin{equation}
\delta I^\mathrm{mut}\approx\delta S_\mathrm{sys}-\frac{\overline{\delta}Q}{k_\mathrm{B}T}=\delta S^\mathrm{irr}=-\mathcal{F}\delta t
\label{eq_infi_small_mut_info}
\end{equation}
up to a negligible high-order term $\mathcal{O}\left(\delta t^2\right)$. Consequently, the information flux quantified by the entropy production rate is shared by the system
and can be used to establish the system-environment correlations. Detailed calculations are shown in Appendix~\ref{app_info_flux_prot}.

\section{LINDBLAD SUPEROPERATOR PRESCRIPTION}

One of the most important approaches in open quantum systems is the well-known Lindblad master equation \cite{gorini_lindbladform_jmp_1976,lindblad_comm_math_phys_1976}. The
dissipative effects caused by the environment are described by the standard Lindblad superoperators acting on system density operator
\begin{equation}
\mathfrak{R}_{i,j}\left\{\rho\right\}=\widehat{A}_i\rho\widehat{A}_j^\dagger-\frac{1}{2}\left\{\widehat{A}_j^\dagger\widehat{A}_i,\rho\right\}.
\end{equation}
Each superoperator $\mathfrak{R}_{i,j}$ is associated with a decay rate $\gamma _{i,j}$. In general, these rates can be time varying. The non-Markovianity and indivisibility
of a dynamical map $\mathcal{E}_t$ is characterized by the Kossakowski matrix $\mathcal{K}_t=\left[\gamma_{i,j}(t)\right]_{(n-1)\times(n-1)}$ formed by collecting the decay
rates. If $\mathcal{K}_t$ is positive semidefinite for all time instances, $\mathcal{E}_t$ are shown to be CP divisible and Markovian. On the other hand, if some eigenvalues
of $\mathcal{K}_t$ temporarily become negative, $\mathcal{E}_t$ then deviates from being CP divisible and exhibits a hierarchy of non-Markovianity. However, the
non-Markovianity usually cannot be detected by quantity-based measures unless $\mathcal{E}_t$ exhibits the essential non-Markovianity
\cite{chruscinski_k_divi_prl_2014,hongbin_k_divi_pra_2015} or strong non-Markovianity\cite{bernardes_experiment_k_divi_scirep_2015}.

Now we are ready to precisely describe the thermodynamic quantities discussed so far within an open system framework. According to the definitions in
Ref.~\cite{alicki_open_sys_th_jpa_1979}, the heat absorption rate by the system is defined as
\begin{equation}
\frac{\partial Q}{\partial t}=\sum_{i,j}\gamma_{i,j}\mathrm{Tr}\left(\mathfrak{R}_{i,j}\left\{\rho\right\}\cdot\widehat{H}_{\mathrm{sys}}\right).
\label{eq_rat_heat_lindblad_form}
\end{equation}
And the changing rate of the system entropy is
\begin{equation}
\frac{\partial S_\mathrm{sys}}{\partial t}=\sum_{i,j}-\gamma_{i,j}\mathrm{Tr}\left(\mathfrak{R}_{i,j}\left\{\rho\right\}\cdot\mathrm{ln}\rho\right).
\label{eq_rat_entr_lindblad_form}
\end{equation}
Combining Eqs.~(\ref{eq_rat_heat_lindblad_form}) and (\ref{eq_rat_entr_lindblad_form}), the entropy production rate in the Lindblad prescription is
given by
\begin{equation}
\frac{\partial S^\mathrm{irr}}{\partial t}=\sum_{i,j}-\gamma_{i,j}\mathrm{Tr}\left(\mathfrak{R}_{i,j}\left\{\rho\right\}\cdot
\left(\mathrm{ln}\rho+\frac{1}{k_\mathrm{B}T}\widehat{H}_\mathrm{sys}\right)\right).
\label{eq_rat_entr_prod_lindblad_form}
\end{equation}
If we image each superoperator $\mathfrak{R}_{i,j}$ defines an interaction channel with the environment, according to definition~(\ref{eq_info_flux_definition}) and
Eq.~(\ref{eq_rat_entr_prod_lindblad_form}), the total information flux can be written as a summation over the flux through each interaction channel
$\mathcal{F}=\sum_{i,j}\mathcal{F}_{i,j}$, where
\begin{equation}
\mathcal{F}_{i,j}=\gamma_{i,j}\mathrm{Tr}\left(\mathfrak{R}_{i,j}\left\{\rho\right\}\cdot\left(\mathrm{ln}\rho+\frac{1}{k_\mathrm{B}T_{i,j}}\widehat{H}_\mathrm{sys}\right)\right).
\label{eq_rat_info_flu_lindblad_form}
\end{equation}
The right-hand-side of Eq.~(\ref{eq_rat_info_flu_lindblad_form}) is proportional to the decay rate $\gamma_{i,j}$ and it therefore concludes one of our main results,
connecting the information flux with the non-Markovianity of system dynamics.

\section{HIERARCHY OF NON-MARKOVIANITY}

For textual completeness and convenience in the following discussions, here we briefly review the concepts of $k$ positivity and hierarchy of non-Markovianity
\cite{chruscinski_k_divi_prl_2014,hongbin_k_divi_pra_2015}. Let $\mathcal{C}$ be a C$^{\ast}$ algebra of linear operators on the $n$-dimensional Hilbert space
$\mathcal{H}_n$, $\mathcal{C}^+$ be the subset of positive elements in $\mathcal{C}$, and $\mathfrak{L}\left(\mathcal{C},\mathcal{C}\right) $ denote the set of linear maps
from $\mathcal{C}$ to $\mathcal{C}$. A TP map $\mathcal{E}\in\mathfrak{L}\left(\mathcal{C},\mathcal{C}\right)$ is said to be positive if
$\mathcal{E}\left(\mathcal{C}^+\right)\subseteq\mathcal{C}^+$. Namely, $\mathcal{E}$ preserves the positivity of the domain $\mathcal{C}^+$.

Since a quantum system may be entangled with some other ancillary degrees of freedom, the notion of positivity of a map should be generalized to $k$ positivity to ensure
the validity of the map in the presence of entanglement. A TP map $\mathcal{E}\in\mathfrak{L}\left(\mathcal{C},\mathcal{C}\right)$ is said to be $k$ positive if
$\mathcal{I}_k\otimes\mathcal{E}:\mathcal{M}_k\otimes\mathcal{C}\rightarrow \mathcal{M}_k\otimes\mathcal{C}$ is positive and CP if $\mathcal{E}$ is $k$ positive for all
positive integers $k$, where $\mathcal{I}_{k}$ is the identity map acting on the $k\times k$ matrix algebra $\mathcal{M}_k$.

Having the notion of $k$ positivity, we can generalize CP divisibility to a hierarchy of $k$ divisibility: an invertible CPTP dynamical process
$\mathcal{E}_{t,0}$ is said to be $k$ divisible if, $\forall~t,\tau > 0$, the complement process
\begin{equation}
\Lambda_{t+\tau,t}=\mathcal{E}_{t+\tau,0}\circ\left[\mathcal{E}_{t,0}\right]^{-1}
\end{equation}
is $k$ positive. Accordingly, $n$ divisibility is equivalent to CP divisibility and $\mathcal{E}_{t,0}$ is zero divisible if $\Lambda_{t+\tau,t}$ violates the positivity
for some $t$ or $\tau$. Introducing a family of sets $\mathcal{D}_k$ containing processes $\mathcal{E}_{t,0}$ with divisibility less than $k$, one has a chain of
inclusions,
\begin{equation}
\mathcal{D}_0\subseteq\mathcal{D}_1\subseteq\cdots\subseteq\mathcal{D}_{n-1}\subseteq\mathcal{D}_n,
\label{k_divi_inclusion_chain}
\end{equation}
where $\mathcal{D}_n$ consists of all CPTP dynamical processes, regardless of their degree of divisibility. In particular, $\mathcal{D}_0$ consists of zero-divisible
processes, which is said to be essentially non-Markovian \cite{chruscinski_k_divi_prl_2014,hongbin_k_divi_pra_2015} or strong non-Markovian
\cite{bernardes_experiment_k_divi_scirep_2015}, and all processes in $\mathcal{D}_{n-1}-\mathcal{D}_0$ are said to be weakly non-Markovian
\cite{chruscinski_k_divi_prl_2014,hongbin_k_divi_pra_2015,bernardes_experiment_k_divi_scirep_2015}. We can further define the sets of proper $k$ divisibility
$\mathcal{PD}_k=\mathcal{D}_k-\mathcal{D}_{k-1}$; then $\mathcal{PD}_n=\mathcal{D}_n-\mathcal{D}_{n-1}$ consists of processes which are exactly $n$ divisible (i.e.,
CP divisible), and therefore Markovian, processes, and $\mathcal{PD}_0=\mathcal{D}_0$. Thus the above inclusion chain can be expressed as a partition of $\mathcal{D}_n$
in terms of $\mathcal{PD}_k$
\begin{equation}
\mathcal{D}_n=\bigcup_{k=0}^n\mathcal{PD}_k.
\label{k_divi_partition}
\end{equation}
It is therefore convenient to visualize the partition in Eq.~(\ref{k_divi_partition}) in terms of a $k$-divisibility phase diagram \cite{hongbin_k_divi_pra_2015} and
investigate the dependence of $k$ divisibility on dynamical parameters of interest.

\section{CNOT GATE}

\subsection{Dynamics of T qubit}

As an instructive paradigm, we consider a pair of qubits coupled with each other via a CNOT gate. The initial state of the control (C) qubit is assumed to be a mixture
$\rho_\mathrm{C}=a|1_\mathrm{C}\rangle \langle 1_\mathrm{C}|+(1-a)|0_\mathrm{C}\rangle \langle 0_\mathrm{C}|$ with $a\in[0,1]$. The qubit pair has no initial interqubit
correlation and their interaction can be described by the Hamiltonian
$\widehat{H}_\mathrm{CT}=\frac{J}{2}\left(|1_\mathrm{C}\rangle\langle 1_\mathrm{C}|\otimes\hat{\sigma}_x+|0_\mathrm{C}\rangle \langle 0_\mathrm{C}|\otimes\widehat{\mathcal{I}}\right)$.
In addition, we impose noisy isotropic depolarizing channels on the target (T) qubit. Although the entire dynamics of the qubit pair is Markovian, it is not the case if
we consider the dynamics of the T qubit after tracing out the C qubit. It is governed by the master equation
\begin{eqnarray}
\frac{\partial}{\partial t}\rho_\mathrm{T}&=&-\frac{i}{\hbar}\left[
\mathcal{J}_x(t)\hat{\sigma}_x,\rho_\mathrm{T}\right]+\frac{\gamma+\gamma_{\mathrm{C},x}(t)}{2}\left(\hat{\sigma}_x\rho_\mathrm{T}\hat{\sigma}_x-\rho_\mathrm{T}\right)  \nonumber \\
&&+\sum_{j=y,z}\frac{\gamma}{2}\left(\hat{\sigma}_j\rho_\mathrm{T}\hat{\sigma}_j-\rho_\mathrm{T}\right) ,
\label{eq_master_equation_t_qubit}
\end{eqnarray}
where $\mathcal{J}_x(t)=\left(J/2r^{2}(t)\right)\left(a^{2}+a(1-a)\cos(Jt/\hbar)\right)$,
\begin{equation}
\gamma_{\mathrm{C},x}(t)=\frac{a(1-a)J}{\hbar r^2(t)}\sin\frac{J}{\hbar}t,
\end{equation}
and $r(t)=\sqrt{(1-a)^{2}+2a(1-a)\cos(Jt/\hbar)+a^{2}}$. Further detailed solutions can be found in Appendix~\ref{app_info_flux_cnot}.

In this paradigm, the T qubit couples to two environments. One is the Markovian isotropic depolarizing channels, which attempts to wash out all information in the T qubit
and push it toward a completely mixed state. Hence the corresponding temperature is assumed to be infinitely high in accordance with the notion of virtual temperature
\cite{skrzypczyk_work_ext_nat_comm_2014,brunner_virt_qubit_virt_temp_pre_2012}. The other environment is played by the C qubit, which introduces non-Markovianity into the
T qubit dynamics in terms of the time-varying rate $\gamma_{\mathrm{C},x}(t)$ associated to $\hat{\sigma}_x$ channel. It is interesting to notice that the C qubit
consists of only two states, far from being an authentic reservoir. However, our definitions~(\ref{eq_info_flux_definition}) and (\ref{eq_rat_info_flu_lindblad_form})
still hold since the C qubit has a static population during the entire dynamics and therefore behaves as a ``static environment."

\subsection{$k$ divisibility and retrieved information}

\begin{figure*}[tbp]
\includegraphics[width=\textwidth]{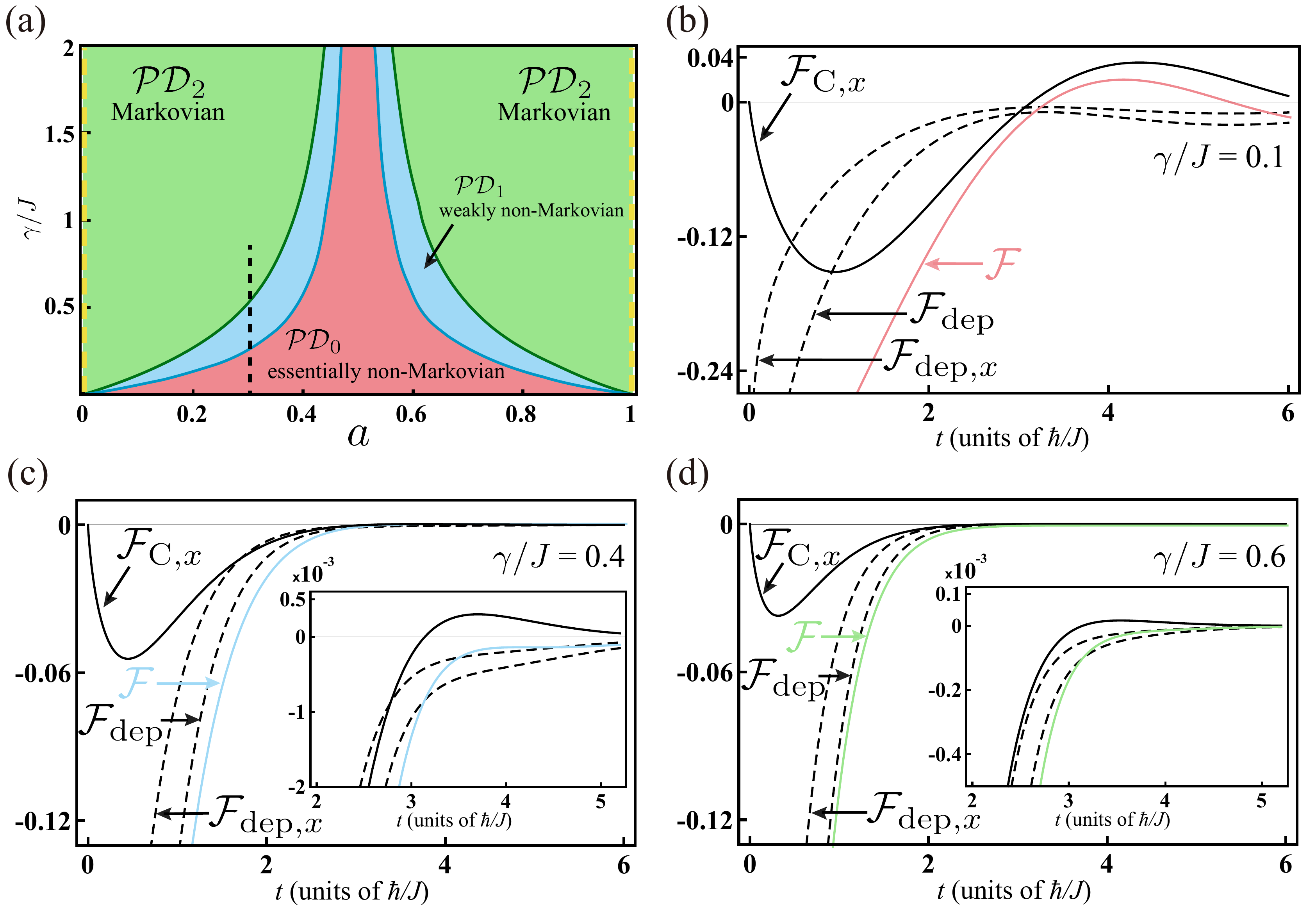}
\caption{(a) k-divisibility phase diagram for the T qubit dynamics vs $a$ and $\gamma/J$. It is CP divisible and Markovian for $a=0,1$, as indicated by the two yellow
dashed lines in the green $\mathcal{PD}_2$ region. As $a$ approaches $0.5$ or $\gamma$ decreases, the T qubit dynamics shows transition from $\mathcal{PD}_2$
to $\mathcal{PD}_0$ region, and therefore shows enhancements in indivisibility and non-Markovianity. The black dashed line represents the case for $a=0.3$,
along which the numerical results and the BLP measure are depicted following and in Fig.~\ref{fig_blp_measure_cnot}, respectively.
In panels (b), (c), and (d), we show the detailed information flux for the T qubit. In these calculations we fix $a=0.3$ and increase
$\gamma/J=0.1$, $0.4$, to $0.6$, respectively. The information flux induced by the C qubit via $\hat{\sigma}_x$ channel, $\mathcal{F}_{\mathrm{C},x}$
(black solid curve in each panel), shows temporary positive periods. This reveals the competitions with dephasings via $\hat{\sigma}_x$ channel,
$\mathcal{F}_{\mathrm{dep},x}$, or the whole depolarizing channels, $\mathcal{F}_\mathrm{dep}$ (black dashed curves). The resulting total information
flux, $\mathcal{F}$ (colored curve in each panel), shows different behavior in each panel, corresponding to three regions in the $k$-divisibility phase
diagram.}
\label{fig_results_cnot}
\end{figure*}

The non-Markovian features of the T qubit are shown in the $k$-divisibility phase diagram Fig.~\ref{fig_results_cnot}(a). If $\gamma_{\mathrm{C},x}(t)=0$ for all
$t$ (e.g., $a=0,1$), this corresponds to the two yellow dashed lines in the green Markovian $\mathcal{PD}_2$ region. Namely the T qubit experiences the
Markovian evolution and the positive decay rate $\gamma$ implies that the information is continuously washed out due to three depolarizing channels. As
$a$ approaches $0.5$ or $\gamma$ decreases, the T qubit dynamics shows transition from Markovian $\mathcal{PD}_2$ to essentially non-Markovian
$\mathcal{PD}_0$ region and therefore exhibits non-Markovianity and indivisibility.

This landscape of non-Markovianity is a result of the competition between the retrieved information and dephasing. If the amplitude of $\gamma_{\mathrm{C},x}(t)$
is finite, its oscillating behavior implies that partial information is periodically flowing out of the T qubit and is subsequently retrieved from the
correlations with C qubit. The numerical results are shown in Figs.~\ref{fig_results_cnot}(b)-\ref{fig_results_cnot}(d), corresponding to the black dashed line at $a=0.3$ in
Fig.~\ref{fig_results_cnot}(a).

In Fig.~\ref{fig_results_cnot}(b), we assume a small $\gamma/J$ value at $0.1$ such that the amplitude of $\gamma_{\mathrm{C},x}(t)$ is larger than $2\gamma$. The
information flux induced by the C qubit via $\hat{\sigma}_x$ channel, $\mathcal{F}_{\mathrm{C},x}$ (black solid curve), becomes temporarily positive
after an initial negative period, revealing substantial retrieved information which overcomes not only the dephasing via the depolarizing
$\hat{\sigma}_x$ channel, $\mathcal{F}_{\mathrm{dep},x}$ (black dashed curve), but also three isotropic depolarizing channels,
$\mathcal{F}_\mathrm{dep}=\sum_{j=x,y,z}\mathcal{F}_{\mathrm{dep},j}$ (black dashed curve). This competition results in the positive periods of total information flux,
$\mathcal{F}=\mathcal{F}_{\mathrm{C},x}+\mathcal{F}_\mathrm{dep}$ (red solid curve). This explicit backflow of information can be detected by the BLP measure
\cite{BLP_measure_prl_2009} and the T qubit dynamics is essentially non-Markovian, zero divisible, and corresponds to the red $\mathcal{PD}_0$ region in Fig.~\ref{fig_results_cnot}(a).

If the amplitude of $\gamma_{\mathrm{C},x}(t)$ lies between $\gamma$ and $2\gamma$, as shown in Fig.~\ref{fig_results_cnot}(c) with $\gamma/J=0.4$, the
transiently retrieved information, $\mathcal{F}_{\mathrm{C},x}$ (black solid curve), within the periods of positive values, is possible to overcome the
dephasing via the $\hat{\sigma}_x$ channel, $\mathcal{F}_{\mathrm{dep},x}$ (black dashed curve); more precisely, there exists some time periods such that
$\mathcal{F}_x=\mathcal{F}_{\mathrm{C},x}+\mathcal{F}_{\mathrm{dep},x}>0$. Therefore, the T qubit can temporarily receive the retrieved information via
$\hat{\sigma}_x$ channel and its dynamics shows weak non-Markovianity, 1 divisibility, and deviating from being CP divisible. However, even though the T
qubit can receive the temporarily retrieved information via $\hat{\sigma}_x$ channel, it is not strong enough and will be
smeared by the $\hat{\sigma}_y$ and $\hat{\sigma}_z$ depolarizing channels. Consequently, the total information flux,
$\mathcal{F}=\mathcal{F}_x+\mathcal{F}_{\mathrm{dep},y}+\mathcal{F}_{\mathrm{dep},z}$ (blue solid curve), is negative
and the quantity-based measures end up with null non-Markovianity in the blue $\mathcal{PD}_1$ region in Fig~\ref{fig_results_cnot}(a) \cite{hongbin_k_divi_pra_2015}.

Figure~\ref{fig_results_cnot}(d) shows the result of $\gamma/J=0.6$. In this case, $|\gamma_{\mathrm{C},x}(t)|\leq\gamma$ for all $t$. The transiently retrieved information
is too weak to compensate the dephasing via $\hat{\sigma}_x$ channels. Hence both $\mathcal{F}_x$ and the total information flux $\mathcal{F}$ are flowing out of the T
qubit and its dynamics is CP divisible and Markovian, corresponding to the green $\mathcal{PD}_2$ region in Fig.~\ref{fig_results_cnot}(a).

\subsection{Non-Markovianity and retrieved information}

\begin{figure}[th]
\includegraphics[width=\columnwidth]{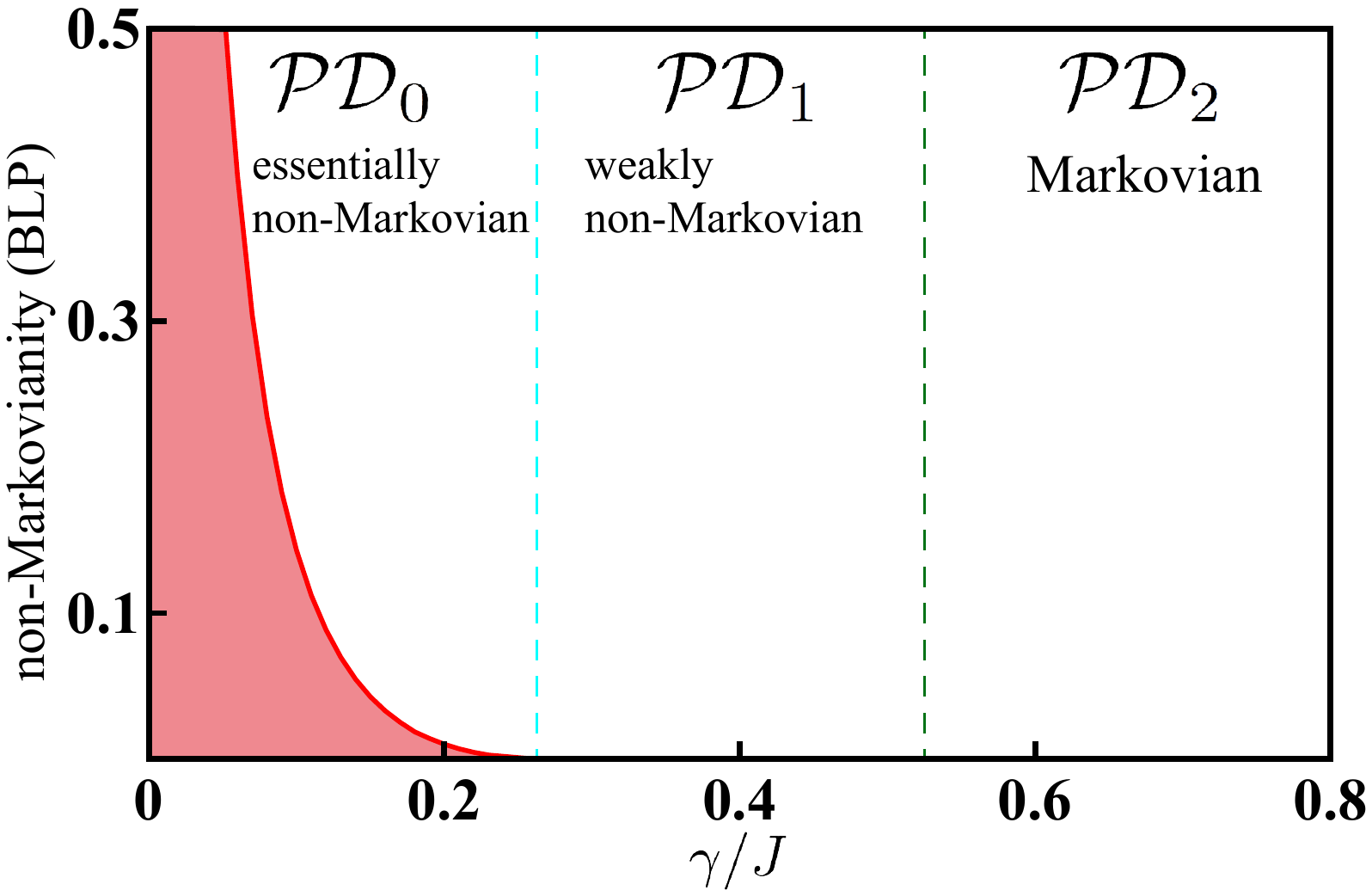}
\caption{The BLP measure decreases rapidly with increasing $\gamma$. The measure can only detect the non-Markovianity in the zero-divisible $\mathcal{PD}_0$ region since the
information backflow is strong enough to increase the trace distance. In the $\mathcal{PD}_1$ region, although partial information can also be retrieved, it is not enough
to overcome the dephasing by the isotropic depolarizing channels. The overall information is outgoing and therefore the BLP measure is blind to this partial information
backflow.}
\label{fig_blp_measure_cnot}
\end{figure}

To further reveal the connection between the quantity-based measures and information backflow, in Fig.~\ref{fig_blp_measure_cnot} we show the BLP measure
\cite{BLP_measure_prl_2009} along the black dashed line at $a=0.3$ in Fig.~\ref{fig_results_cnot}(a). The BLP measure decreases rapidly with increasing $\gamma$ and
identifies nonzero non-Markovianity only in the $\mathcal{PD}_0$ region. This can be understood from Fig.~\ref{fig_results_cnot}(b), where the total information flux
$\mathcal{F}$ shows positive periods, revealing strong enough information backflow resulting in the increments of trace distance and nonzero BLP measure.

The equivalence between total information flux $\mathcal{F}$ and BLP measure can be realized by observing that $\mathcal{F}$ is proportional to the time varying rate
of trace distance of a specific state pair:
\begin{equation}
\mathcal{F}=\sum_{j=x,y,z}-\frac{\partial S_j}{\partial t}
\propto\frac{\partial}{\partial t}\frac{1}{2}
\Vert\mathcal{E}_t\left(|1\rangle\langle 1|-|0\rangle\langle 0|\right)\Vert_1,
\end{equation}
where $\mathcal{E}_t$ is the process generated by the master equation~(\ref{eq_master_equation_t_qubit}). The trace distance revives only when the total information is
incoming, no matter how much detailed information through each interaction channel is transiently retrieved.

It is worthwhile to notice that, in the $\mathcal{PD}_1$ region, partial information can be retrieved via $\hat{\sigma}_x$ channel, resulting in a low degree of
indivisibility (i.e., 1 divisibility), whereas the quantity-based measures tend to be blind to the weak non-Markovianity in the $\mathcal{PD}_1$ region
\cite{hongbin_k_divi_pra_2015}. This is because they can at most detect the total information flux $\mathcal{F}$, which is outgoing in the $\mathcal{PD}_1$ region, rather than
access the detailed flux through each interaction channel, which is possibly incoming, as seen from Fig.~\ref{fig_results_cnot}(c). In other words, if solely relying on the
quantity-based measures, one can neither detect the weak non-Markovianity nor distinguish between $\mathcal{PD}_1$ and $\mathcal{PD}_2$. This can only be understood when
one analyzes the detailed information flux through each interaction channel.

\section{DISCUSSIONS AND CONCLUSIONS}

Finally, we explore the possibility of experimental implementation of our protocol presented above. We briefly discuss
two types of promising candidates. We notice that the interaction Hamiltonian~(\ref{eq_prot_int_hami}) is of the form of the Jaynes-Cummings model within the rotating-wave
approximation. It has been shown that the linear optical setups are competent for simulating such systems and achieving several thermodynamic tasks
\cite{guzik_review_photo_q_simu_nat_phys_2012,jinshi_alg_q_cooling_nat_photo_2014,luca_qubit_thermometry_prl_2017}. Additionally, they can be used to demonstrate the transition
between different non-Markovian regimes as well \cite{bernardes_experiment_k_divi_scirep_2015,BLP_measure_exp_nat_phys_2011}. On the other hand, thanks to the massive efforts
devoted to the studies of nanoscale devices, considerable improvements in the fabrication and the manipulation of the electronic circuits have been realized. Many experiments
have been performed for verifying the fundamental theories of classical and quantum thermodynamics \cite{jukka_review_electronic_qtd_nat_phys_2015,koski_maxwell_demon_exp_prl_2015}.
Based on the experiments reported above, we believe that our approach may have potential applications in various types of quantum heat engines
\cite{scully_qhe_pnas_2011,rossnage_1_atom_qhe_science_2016,hongbin_vib_coh_bio_qhe_2016,sachtleben_supercond_qhe_prl_2017}.

In summary, our main results exhibited that, when a system interacts with a static environment, the information flux is equal to the negative entropy production rate. The
system attempts to share this outflowing information with the environment and establish system-environment correlations. For these results, we revisited the thermodynamic
task of work extraction and the second law of thermodynamics. We quantified the amount of information in a system by the relative entropy with respect to its thermal
equilibrium and described how this information changes during a dynamical process. We further presented a simple protocol to reaffirm our arguments.

Invoking the Lindblad superoperator prescription, we investigated the information flux within the framework of open system. We found that the indivisibility of the dynamics
is intimately connected to the direction of information flux. In general, a higher degree of non-Markovianity or indivisibility implies a stronger backflow of information.
To explicitly reveal the connection between non-Markovianity and information backflow, we considered the CNOT gate model. We found that when increasing the strength of
information backflow, the dynamics of the target qubit transfers from being Markovian to non-Markovian and shows a higher degree of non-Markovianity and indivisibility. This
supports the physical interpretation of the BLP measure of non-Markovianity and shows that the quantity based measures
\cite{BLP_measure_prl_2009,RHP_measure_prl_2010,LFS_measure_pra_2012,LPP_measure_pra_2013,BCM_measure_sci_rep_2014,acc_info_measure_prl_2014,acc_info_measure_pra_2014,shinlian_tsw_measure_prl_2016}
are not sensitive enough to capture the detailed information backflow.

($\equiv\widehat{\mathsf{\Phi}}\omega\widehat{\mathsf{\Phi}}\equiv$)$\sim$meow$\sim$

\section*{ACKNOWLEDGMENTS}

We acknowledge Neill Lambert, Ken Funo, and Philipp Strasberg for helpful discussions and feedback. This work is supported partially by the National Center for Theoretical
Sciences and Ministry of Science and Technology, Taiwan, Grants No. MOST 103-2112-M-006-017-MY4, No. MOST 105-2811-M-006-059, and No. MOST 105-2112-M-005-008-MY3.

\appendix

\section{DERIVATION OF EQ.~(\ref{eq_entr_pro_rel_entr})} \label{app_derivation_eq_entr_pro_rel_entr}

Here we show the detailed derivation of Eq.~(\ref{eq_entr_pro_rel_entr}). The approach is similar to that in Ref.~\cite{esposito_ent_prod_corr_njp_2010}. As mentioned in the
main context, we do not assume thermalized nor static environment in the beginning of our derivation. The environment is kept general. Since the system and environment are
considered in totality as closed, the total state evolves unitarily without a change in total entropy. Besides, the initial total state is assumed to be a direct product of
system and environment; we therefore have
\begin{equation}
S_\mathrm{tot}(t)=S_\mathrm{tot}(0)=S_\mathrm{sys}(0)+S_\mathrm{env}(0).
\end{equation}
And the change in system entropy can be written as
\begin{eqnarray}
\Delta S_\mathrm{sys}(t)&=&S_\mathrm{sys}(t)-S_\mathrm{sys}(0) \nonumber\\
&=&-\mathrm{Tr}\rho_\mathrm{sys}(t)\mathrm{ln}\rho_\mathrm{sys}(t) \nonumber\\
&&+\mathrm{Tr}\rho_\mathrm{tot}(t)\mathrm{ln}\rho_\mathrm{tot}(t)-\mathrm{Tr}\rho_\mathrm{env}(0)\mathrm{ln}\rho_\mathrm{env}(0). \nonumber\\
\end{eqnarray}

By noticing that $\mathrm{Tr}\rho_\mathrm{sys}(t)\mathrm{ln}\rho_\mathrm{sys}(t)=\mathrm{Tr}\rho_\mathrm{tot}(t)\mathrm{ln}\left[\rho_\mathrm{sys}(t)\otimes\widehat{\mathcal{I}}_\mathrm{env}\right]$,
where $\widehat{\mathcal{I}}_\mathrm{env}$ is the identity operator acting on the environmental Hilbert space, we have
\begin{eqnarray}
\Delta S_\mathrm{sys}(t)&=&S\left(\rho_\mathrm{tot}(t)\Vert\rho_\mathrm{sys}(t)\otimes\rho_\mathrm{env}^\mathrm{eq}\right) \nonumber\\
&&-\mathrm{Tr}\rho_\mathrm{env}(0)\mathrm{ln}\rho_\mathrm{env}(0)+\mathrm{Tr}\rho_\mathrm{env}(t)\mathrm{ln}\rho_\mathrm{env}^\mathrm{eq} \nonumber\\
&=&S\left(\rho_\mathrm{tot}(t)\Vert\rho_\mathrm{sys}(t)\otimes\rho_\mathrm{env}^\mathrm{eq}\right)
-S\left(\rho_\mathrm{env}(0)\Vert\rho_\mathrm{env}^\mathrm{eq}\right) \nonumber\\
&&+\mathrm{Tr}\left[\rho_\mathrm{env}(t)-\rho_\mathrm{env}(0)\right]\mathrm{ln}\rho_\mathrm{env}^\mathrm{eq}.
\end{eqnarray}
Due to the closure of the total system, the environment can only exchange heat with the system. The last term is equal to $\overline{\Delta}Q/k_\mathrm{B}T$.
We obtain the first line of Eq.~(\ref{eq_entr_pro_rel_entr}) that
\begin{eqnarray}
\Delta S^\mathrm{irr}&=&S\left(\rho_\mathrm{tot}(t)\Vert\rho(t)\otimes\rho_\mathrm{env}^\mathrm{eq}\right)
-S\left(\rho_\mathrm{env}(0)\Vert\rho_\mathrm{env}^\mathrm{eq}\right). \nonumber \\
\end{eqnarray}

We proceed to expand the first relative entropy on the right-hand side of Eq.~(\ref{eq_entr_pro_rel_entr}). Simple algebraic skill leads to
\begin{eqnarray}
S\left(\rho_\mathrm{tot}(t)\Vert\rho(t)\otimes\rho_\mathrm{env}^\mathrm{eq}\right)&=&S_\mathrm{sys}(t)-S_\mathrm{tot}(t) \nonumber\\
&&-\mathrm{Tr}\rho_\mathrm{env}(t)\mathrm{ln}\rho_\mathrm{env}^\mathrm{eq} \nonumber\\
&&+S_\mathrm{env}(t)+\mathrm{Tr}\rho_\mathrm{env}(t)\mathrm{ln}\rho_\mathrm{env}(t). \nonumber\\
\end{eqnarray}
We finally obtain the second line of Eq.~(\ref{eq_entr_pro_rel_entr}) that
\begin{equation}
\Delta S^\mathrm{irr}=I^\mathrm{mut}(t)+I_\mathrm{env}^\mathrm{neq}(t)-I_\mathrm{env}^\mathrm{neq}(0).
\end{equation}

It is interesting to notice that, if we adopt a thermalized initial environment state $\rho_\mathrm{env}(0)=\rho_\mathrm{env}^\mathrm{eq}$,
we have
\begin{equation}
I_\mathrm{env}^\mathrm{neq}(0)=S\left(\rho_\mathrm{env}(0)\Vert\rho_\mathrm{env}^\mathrm{eq}\right)=0.
\end{equation}
We therefore recover the results in Eq.~(\ref{eq_entr_pro_rel_entr}). Alternatively, if we assume the static environment hypothesis,
then we have $I_\mathrm{env}^\mathrm{neq}(t)=I_\mathrm{env}^\mathrm{neq}(0)$. Consequently entropy production reduces to system-environment correlations
exclusively
\begin{equation}
\Delta S^\mathrm{irr}=I^\mathrm{mut}(t).
\end{equation}
Therefore, taking the time derivative form of the above equation recovers our main result in Eq.~(\ref{eq_rat_mut_info_neg_info_flux}) immediately.

\section{INFORMATION FLUX OF THE PROTOCOL} \label{app_info_flux_prot}

Here we show further details regarding our protocol. The notion of virtual qubit
\cite{skrzypczyk_work_ext_nat_comm_2014,brunner_virt_qubit_virt_temp_pre_2012} is one of the critical ingredients in our protocol. Whenever we specify
certain two states of the environment as a virtual qubit on resonance with the system, then the state of the environment can be expressed as
\begin{equation}
\rho_\mathrm{env}=\left(\frac{q_1}{Z}|1\rangle\langle 1|+\frac{q_0}{Z}%
|0\rangle\langle 0|\right)\oplus\rho^\mathrm{red}_\mathrm{env},
\end{equation}
where $q_1/q_0=\exp\left[-(E_a-E_b)/k_\mathrm{B}T\right]$, and $\rho^\mathrm{red}_\mathrm{env}$ is the redundant state apart from the virtual qubit with
$\mathrm{Tr}\rho^\mathrm{red}_\mathrm{env}=1-q_1/Z-q_0/Z$.

The initial state $\rho_\mathrm{tot}(0)$ is a direct product of system and environment. The infinitesimal evolution of the system and virtual qubit in the
first stage can be written as
\begin{eqnarray}  \label{eq_joi_den_mat_sys_vir_qub}
&&\rho_\mathrm{sys}\otimes\rho_\mathrm{vir}=\left[
\begin{array}{cc}
p_a & 0 \\
0 & p_b
\end{array}
\right] \otimes \frac{1}{Z}\left[
\begin{array}{cc}
q_1 & 0 \\
0 & q_0
\end{array}
\right]  \nonumber \\
&&\rightarrow \left[
\begin{array}{cccc}
p_a\frac{q_1}{Z} & 0 & 0 & 0 \\
0 & p_a\frac{q_0}{Z}-p_Z\delta t & ip_Z\delta t & 0 \\
0 & -ip_Z\delta t & p_b\frac{q_1}{Z}+p_Z\delta t & 0 \\
0 & 0 & 0 & p_b\frac{q_0}{Z}
\end{array}
\right]+\mathcal{O}\left(p_Z^2\delta t^2\right),  \nonumber \\
\end{eqnarray}
where $p_Z=\left(p_aq_0-p_bq_1\right)\gamma/Z$. The off-diagonal elements reveal that a non-classical correlation is established during the infinitesimal
evolution in this stage.

In the second stage, we erase the system-environment correlation and obtain the reduced density matrices for the system and environment:
\begin{equation}
\rho_\mathrm{sys}(\delta t)=\left[
\begin{array}{cc}
p_a-p_Z\delta t & 0 \\
0 & p_b+p_Z\delta t
\end{array}
\right],  \label{eq_den_mat_sys_dt}
\end{equation}
\begin{equation}
\rho_\mathrm{env}(\delta t)=\left[
\begin{array}{cc}
\frac{q_1}{Z}+p_Z\delta t & 0 \\
0 & \frac{q_0}{Z}-p_Z\delta t
\end{array}
\right] \oplus\rho^\mathrm{red}_\mathrm{env}.  \label{eq_den_mat_env_dt}
\end{equation}
The heat absorbed by the environment is equal to the one transferred from the system
\begin{equation}
\overline{\delta}Q=-(E_a-E_b)p_Z\delta t.
\end{equation}
The small change in the entropy of the system is
\begin{equation}
\delta S_\mathrm{sys}=-\mathrm{Tr}\left(\delta\rho_\mathrm{sys}\cdot\mathrm{ln}\rho_\mathrm{sys}\right)=\left(\mathrm{ln}\frac{p_a}{p_b}\right)p_Z\delta t.
\end{equation}
And the one of the environment is
\begin{equation}
\delta S_\mathrm{env}=-\left(\mathrm{ln}\frac{q_1}{q_0}\right)p_Z\delta t=-\frac{\overline{\delta}Q}{k_\mathrm{B}T}.
\end{equation}
This satisfies the results in Eq.~(\ref{eq_diff_entr_env}) that the entropy change in an authentic reservoir solely arises from exchange of heat.

Now we proceed to the quantification of correlation proposed in Ref.~\cite{modi_quan_corr_prl_2010}. Since the initial total state is a direct product
of system and environment, the increment in the correlation is therefore quantified by
$S\left(\rho_\mathrm{tot}(\delta t)\Vert\rho_\mathrm{sys}(\delta t)\otimes\rho_\mathrm{env}(\delta t)\right)$. Modi \textit{et al.}
\cite{modi_quan_corr_prl_2010} have shown that this is equal to the increment in bipartite mutual information $\delta I^\mathrm{mut}$. The relative
entropy can be expanded as
\begin{eqnarray}
&S&\left(\rho_\mathrm{tot}(\delta t)\Vert\rho_\mathrm{sys}(\delta t)\otimes\rho_\mathrm{env}(\delta t)\right)  \nonumber \\
&=&\mathrm{Tr}\rho_\mathrm{tot}(0)\mathrm{ln}\rho_\mathrm{tot}(0) -\mathrm{Tr}\rho_\mathrm{tot}(\delta t)\mathrm{ln}\rho_\mathrm{sys}(\delta
t)\otimes\rho_\mathrm{env}(\delta t)  \nonumber \\
&=&\mathrm{Tr}\rho_\mathrm{sys}\otimes\rho_\mathrm{vir}\mathrm{ln}\rho_\mathrm{sys}\otimes\rho_\mathrm{vir} +\mathrm{Tr}\rho_\mathrm{sys}
\otimes\rho^\mathrm{red}_\mathrm{env}\mathrm{ln}\rho_\mathrm{sys}\otimes\rho^\mathrm{red}_\mathrm{env}  \nonumber \\
&&-\mathrm{Tr}\rho_{\mathrm{sys},\mathrm{vir}}(\delta t)\mathrm{ln}\rho_\mathrm{sys}(\delta t)\otimes\rho_\mathrm{vir}(\delta t)  \nonumber \\
&&-\mathrm{Tr}\rho_\mathrm{sys}\otimes\rho^\mathrm{red}_\mathrm{env}\mathrm{ln}\rho_\mathrm{sys}(\delta t)\otimes\rho^\mathrm{red}_\mathrm{env}.
\label{eq_rel_ent_tot_dir_pro}
\end{eqnarray}
In the first equality, we have used the unitarity of total system such that $S_\mathrm{tot}\left(\delta t\right)=S_\mathrm{tot}\left(0\right)$.
Finally, substituting Eqs.~(\ref{eq_joi_den_mat_sys_vir_qub})-(\ref{eq_den_mat_env_dt}) into Eq.~(\ref{eq_rel_ent_tot_dir_pro}), we can recover the
result in Eq.~(\ref{eq_infi_small_mut_info}).

\section{INFORMATION FLUX THROUGH T QUBIT} \label{app_info_flux_cnot}

As shown in Ref.~\cite{hongbin_k_divi_pra_2015}, the dynamics of T qubit can be expressed as
\begin{equation}
\rho_\mathrm{T}(t)=\alpha_0\rho_{11}(t)+\beta_0\rho_{00}(t)+\delta_0\rho_{01}(t)+\delta^{\ast}_0\rho_{10}(t),
\end{equation}
where $\alpha_0$, $\beta_0$, $\delta_0$, and $\delta^{\ast}_0$ denote the
initial condition of the T qubit and
\begin{eqnarray}
\rho_{11}(t)&=&\left[%
\begin{array}{cc}
\frac{1}{2}\left(1+e^{-2\gamma t}\mathcal{A}(t)\right) & i\frac{a}{2}%
e^{-2\gamma t}\sin\frac{Jt}{\hbar} \\
-i\frac{a}{2}e^{-2\gamma t}\sin\frac{Jt}{\hbar} & \frac{1}{2}%
\left(1-e^{-2\gamma t}\mathcal{A}(t)\right)%
\end{array}%
\right], \\
\rho_{00}(t)&=&\left[%
\begin{array}{cc}
\frac{1}{2}\left(1-e^{-2\gamma t}\mathcal{A}(t)\right) & -i\frac{a}{2}%
e^{-2\gamma t}\sin\frac{Jt}{\hbar} \\
i\frac{a}{2}e^{-2\gamma t}\sin\frac{Jt}{\hbar} & \frac{1}{2}%
\left(1+e^{-2\gamma t}\mathcal{A}(t)\right)%
\end{array}%
\right], \\
\rho_{01}(t)&=&\left[%
\begin{array}{cc}
-i\frac{a}{2}e^{-2\gamma t}\sin\frac{Jt}{\hbar} & ae^{-2\gamma t}\sin^2\frac{%
Jt}{2\hbar} \\
\frac{1}{2}e^{-2\gamma t}\left(1+\mathcal{A}(t)\right) & i\frac{a}{2}%
e^{-2\gamma t}\sin\frac{Jt}{\hbar}%
\end{array}%
\right], \\
\rho_{10}(t)&=&\left[%
\begin{array}{cc}
i\frac{a}{2}e^{-2\gamma t}\sin\frac{Jt}{\hbar} & \frac{1}{2}e^{-2\gamma
t}\left(1+\mathcal{A}(t)\right) \\
ae^{-2\gamma t}\sin^2\frac{Jt}{2\hbar} & -i\frac{a}{2}e^{-2\gamma t}\sin%
\frac{Jt}{\hbar}%
\end{array}%
\right],  \label{dynamics_10_example2}
\end{eqnarray}
with $\mathcal{A}(t)=1-a+a\cos\frac{Jt}{\hbar}$. Having acquired the full dynamics of $\rho_\mathrm{T}(t)$, the master equation~(\ref{eq_master_equation_t_qubit})
can be derived following the methods outlined in Ref.~\cite{andersson_kraus_decompo_jmo_2007}.

For symbolic brevity, we parametrize the initial condition by polar
coordinate $(r_{0},\theta _{0},\phi _{0})$ and it evolves to $(r_{0}\exp
(-2\gamma t)\tilde{r}(t),\theta (t),\phi (t))$ at latter time $t$ with $%
\tilde{r}(t)=\sqrt{(\cos ^{2}\theta _{0}+\sin ^{2}\theta _{0}\sin ^{2}\phi
_{0})r^{2}(t)+\sin ^{2}\theta _{0}\cos ^{2}\phi _{0}}$.

According to the definitions (\ref{eq_rat_heat_lindblad_form})-(\ref{eq_rat_info_flu_lindblad_form}), the heat fluxes via each channel are given
by
\begin{eqnarray}
\frac{\partial Q_x}{\partial t}&=&0, \\
\frac{\partial Q_y}{\partial t}&=&\frac{\partial Q_z}{\partial t}=-\gamma
r_0\sin\theta_0\cos\phi_0\mathcal{J}_x(t)e^{-2\gamma t},
\end{eqnarray}
and the entropy changing rates are given by
\begin{eqnarray}
\frac{\partial S_x}{\partial t}&=&\frac{\gamma+\gamma_x(t)}{2}%
\left(1-\sin^2\theta(t)\cos^2\phi(t)\right)\mathcal{B}(t), \\
\frac{\partial S_y}{\partial t}&=&\frac{\gamma}{2}\left(\sin^2\theta(t)%
\cos^2\phi(t)+\cos^2\theta(t)\right)\mathcal{B}(t), \\
\frac{\partial S_z}{\partial t}&=&\frac{\gamma}{2}\sin^2\theta(t)\mathcal{B}%
(t),
\end{eqnarray}
where $\mathcal{B}(t)=2r_0e^{-2\gamma t}\tilde{r}(t)\mathrm{artanh}\left(r_0e^{-2\gamma t}\tilde{r}(t)\right)$,
with $\mathrm{artanh}(z)=\frac{1}{2}\mathrm{ln}\frac{1+z}{1-z}$ the inverse hyperbolic tangent. As discussed in
the main text, the temperature assigned to the $\hat{\sigma}_y$ and $\hat{\sigma}_z$ channels is infinitely high.
Hence the information flux via each channel is exactly equal to the negative entropy changing rate
\begin{equation}
\mathcal{F}_j=-\frac{\partial S_j}{\partial t},
\end{equation}
where $j=x$, $y$, and $z$. And the total information flux is given by
their summation
\begin{equation}
\mathcal{F}=\sum_j\mathcal{F}_j=\frac{\partial r_0e^{-2\gamma t}\tilde{r}(t)}{%
\partial t}\mathrm{artanh}\left(r_0e^{-2\gamma t}\tilde{r}(t)\right).
\end{equation}

Additionally, in the calculations of information flux in Fig.~\ref{fig_results_cnot}, we adopt the initial condition
$\rho_\mathrm{T}(0)=|1\rangle\langle1|$.


%

\end{document}